# Experimental Demonstration of an Electride as a 2D Material

Daniel L. Druffel,[†] Kaci L. Kuntz,[†‡] Adam H. Woomer,[†‡] Francis M. Alcorn,[†] Jun Hu,[†] Carrie L. Donley,[§] and Scott C. Warren[†§*]

[†] Department of Chemistry and [§] Department of Applied Physical Sciences, University of North Carolina at Chapel Hill, Chapel Hill, North Carolina 27599, United States.

**ABSTRACT:** Because of their loosely bound electrons, electrides offer physical properties useful in chemical synthesis and electronics. For these applications and others, nano-sized electrides offer advantages, but to-date no electride has been synthesized as a nanomaterial. We demonstrate experimentally that $Ca_2N$, a layered electride in which layers of atoms are separated by layers of a 2D electron gas (2DEG), can be exfoliated into two-dimensional (2D) nanosheets using liquid exfoliation. The 2D flakes are stable in a nitrogen atmosphere or in select organic solvents for at least one month. Electron microscopy and elemental analysis reveal that the 2D flakes retain the crystal structure and stoichiometry of the parent 3D $Ca_2N$. In addition, the 2D flakes exhibit metallic character and an optical response that agrees with DFT calculations. Together these findings suggest that the 2DEG is preserved in the 2D material. With this work, we bring electrides into the nano-regime and experimentally demonstrate a 2D electride, $Ca_2N$.

## INTRODUCTION

Nearly every ionic solid consists of positively and negatively charged atoms, but in exotic materials called electrides, the negative "ion" is just an electron with no nucleus. The electron's wavefunction occupies the space typically filled by anions like $Cl^-$.[1-4] Materials with anionic electrons can serve as strong reducing agents[5,6] or catalysts for chemical syntheses,[7,8] and have been predicted to serve as low-temperature electron emitters,[9] transparent conductors,[10,11] and battery electrodes.[12] With so much to be discovered about electrides, their exotic physical properties may open avenues to different kinds of energy storage,[12] optoelectronic,[13] and magnetic devices.

Electrides have two known structures: cage structures[14-16] in which anionic electrons are located within zero-dimensional cages, and layered structures,[17,18] in which anionic electrons are found in two-dimensional (2D) planes. In layered electrides, the proximity of the anionic electrons causes them to partially delocalize[19] as a 2D electron gas.[18] The electron gas enables high electrical mobility (160 $cm^2\,V^{-1}s^{-1}$ at room temperature),[18] high carrier concentrations ($1.4 \times 10^{22}\,cm^{-3}$),[18] and rapid electrical transport to the material's surfaces.

The exciting properties of layered electrides have prompted theoretical studies into their lower dimensional forms. For example, a recent report[20] presented density functional theory (DFT) calculations of 2D $Ca_2N$, which suggest that the 2D form could be thermally and mechanically stable and possesses an electron gas on its surface. That result suggests that 2D $Ca_2N$ might be synthesized by exfoliation of 3D $Ca_2N$. Despite the growing interest, to our knowledge, no electride has been synthesized or experimentally studied as a 2D material or nanomaterial.

Here we demonstrate that electride nanomaterials can be produced by exfoliating a layered electride, $Ca_2N$. We have applied the liquid exfoliation methodology, previously used to break apart van der Waals solids,[21] to exfoliate the layered $Ca_2N$ crystal into 2D nanosheets. As we show, the 2D flakes are crystalline, have a stoichiometry of $Ca_2N_{0.99\pm0.01}$, and are stable in a nitrogen atmosphere or in several organic solvents for at least one month. In addition, our photoemission and optical measurements, which are consistent with the electronic band structure calculated by density functional theory (DFT), suggest that the anionic electrons are retained in the 2D flakes.

## RESULTS AND DISCUSSION

We synthesized bulk $Ca_2N$ by a high-temperature reaction between $Ca_3N_2$ and calcium metal as reported previously,[18,22] with details described in the supporting information. Powder X-ray diffraction patterns of our bulk $Ca_2N$ confirm that the samples have the layered anti-$CdCl_2$-type crystal structure (Figure S1) with lattice parameters matching those in the literature.[17] Figure 1a depicts the crystal structure of $Ca_2N$, in which planes of $Ca_6N$ octahedra are separated by a 3.9 Å interlayer gap.[17,18,23] Because of the formal oxidation states of $Ca^{2+}$ and $N^{3-}$, the formula unit has a positive charge and is best represented as $[Ca_2N]^+$. Anionic electrons balance the positive charge of the $[Ca_2N]^+$ layers by occupying the interlayer gap. Using DFT[24,25], we calculated projections of electron density for the highest occupied band (-1.49 eV to $E_F$, the Fermi level) shown in Figure 1b (see SI for details). We provide an electron density profile with respect to the z-axis of the hexagonal unit cell for this band (Figure 1c) and find that the interlayer electron gas consists of *ca.* 0.7 electrons per formula unit. The bands below the highest occupied band do not contribute additional electron density to the interlayer electron gas.

To understand the energetics of exfoliating $Ca_2N$ into nanosheets, we calculated the binding energy between layers as a function of interlayer distance. We found that our binding energy between $Ca_2N$ layers was 396 meV per calcium atom or 1.11 $J/m^2$. These calculations match those performed by others who report a cleavage strength of 1.09 $J/m^2$ for $Ca_2N$.[20] In addition, we find that the binding energy of $Ca_2N$ is only about four times that of graphite (0.31 $J/m^2$), as shown in Figure 1d. Electrostatic interactions between the $[Ca_2N]^+$ and electron gas likely account for the greater binding energy of $Ca_2N$ compared to graphite, a van der Waals solid. These calculations suggest that it may be possible to exfoliate $Ca_2N$ into 2D flakes, although with somewhat greater difficulty than graphite.

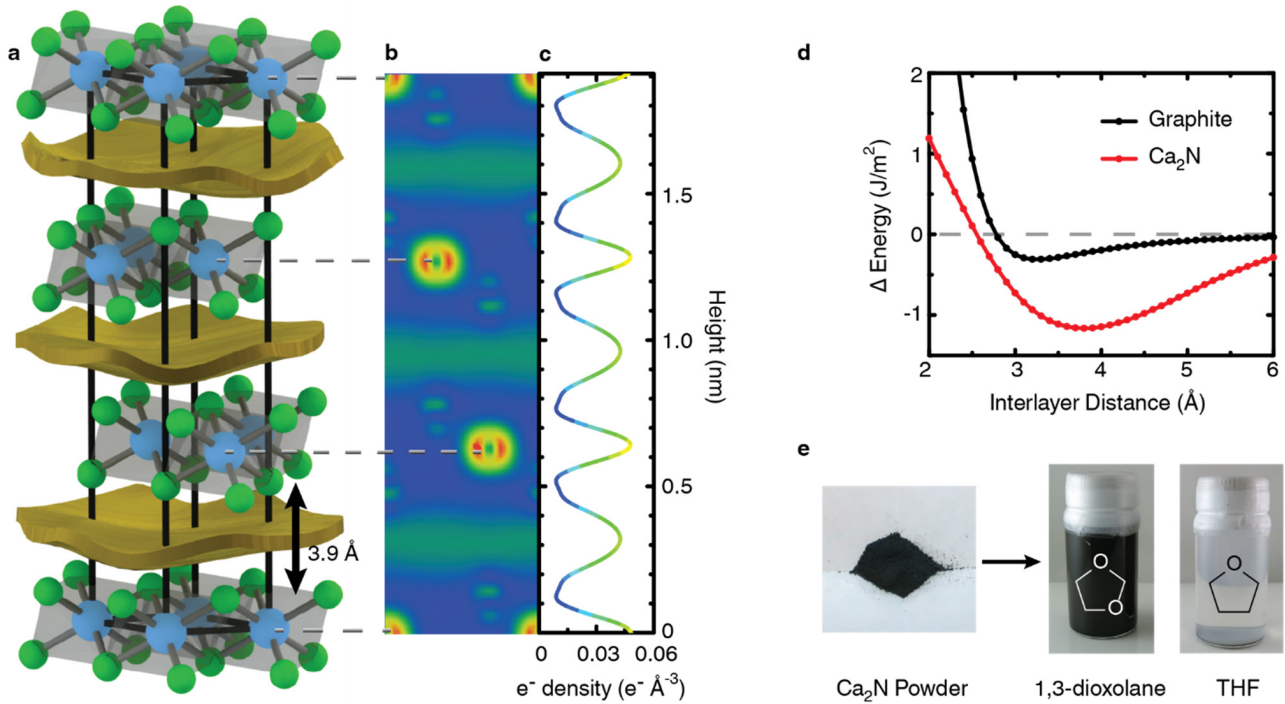

**Figure 1.** Liquid exfoliation of $Ca_2N$. a) The unit cell of $Ca_2N$ depicting layers of $[Ca_2N]^+$ ($Ca^{2+}$ is green, $N^{3-}$ is blue) that alternate with layers of delocalized electrons (gold). b) Projection of the integrated electron density in the unit cell as calculated by density functional theory (DFT). The integration is over occupied states in the highest occupied band. c) The electron density profile integrated along the z-axis of the unit cell for the occupied band shown in b. d) Interlayer binding energy versus interlayer distance, as calculated by DFT. e) Photos of $Ca_2N$ bulk powder and 2D $Ca_2N$ suspensions in 1,3-dioxolane and tetrahydrofuran (THF).

Scotch-tape methods used to exfoliate layered van der Waals solids may not be appropriate for $Ca_2N$. $Ca_2N$ is chemically reactive and can decompose in contact with many adhesives. For such a reactive material, the conditions of liquid exfoliation are more suitable because the solvent's functional groups can be chosen to avoid reaction. Liquid exfoliation offers additional advantages such as a much higher yield, scalability, facile material transfer, and easy thin-film preparation.[21]

To identify solvents that are stable in the presence of $Ca_2N$, we screened 30 solvents (Table S1) with different functional groups. Working under anhydrous and oxygen-free conditions, we suspended bulk $Ca_2N$ powder in each solvent and sonicated the suspensions for 100 minutes in a water bath sonicator (see SI for details).

Reactions between many of the solvents and the bulk $Ca_2N$ were easy to identify by eye. As expected, protic solvents like isopropyl alcohol and N-methylformamide reacted vigorously with $Ca_2N$ to form $Ca(OH)_2$. In other solvents—chloroalkanes, ketones, aldehydes, nitriles, and even some ethers—the dark blue $Ca_2N$ powder decomposed to a white powder in less than 24 hours. Therefore, these solvents are not appropriate for liquid exfoliation. Non-polar hydrocarbons including benzene and hexanes did not react with $Ca_2N$, but the dispersions precipitated within minutes. While non-polar hydrocarbons may not be appropriate for liquid exfoliation, this finding suggests that they could be used to protect $Ca_2N$ from air or water environments. Finally, exfoliation in aprotic amides like N-methyl-2-pyrrolidone showed only modest stability: the materials precipitated rapidly and partially oxidized to $Ca(OH)_2$ over a period of 5 to 10 days.

Through the screening process we identified several promising solvents that did not visibly react with $Ca_2N$, including 1,3-dioxolane, dimethyl carbonate, and dimethoxyethane. These solvents have been successfully employed in lithium-ion batteries[26,27] and are stable against reduction. Though each of these three solvents merits further study, we pursued 1,3-dioxolane because it produced the darkest, most concentrated dispersions and because of its volatility (boiling point = 75 °C).

Although the chemical structures of 1,3-dioxolane, tetrahydrofuran (THF), and 1,4-dioxane are similar, the reactivity of these solvents towards $Ca_2N$ is markedly different. $Ca_2N$ is stable in 1,3-dioxolane for at least a month and remains as a dark suspension, whereas in THF and 1,4-dioxane, it decomposes into white $Ca(OH)_2$ in less than one day (Figure 1e, S2). Further studies are underway to better understand the greater stability of $Ca_2N$ in 1,3-dioxolane.

After identifying 1,3-dioxolane as a promising solvent, we sought to characterize the exfoliated material. Figure 2a shows an image of a nanosheet of $Ca_2N$ acquired using a high-resolution transmission electron microscope (HR-TEM). The material is a single crystal and crystalline out to the edges of the flake, which suggests minimal degradation of the sample during the exfoliation process or while loading into the TEM. The nanosheet is flat and transparent to electrons. Although the predominant morphology is sheet-like, our liquid-exfoliation method does yield some material that is non-sheet-like or that appears to be an aggregate of sheets (Figure S4), despite our attempt to isolate the thinnest materials by centrifugation. We also note that like other liquid exfoliated materials,[21,28,29] our 2D $Ca_2N$ flakes have a distribution of lateral dimensions (approximately 0.2-10 μm) and thicknesses.

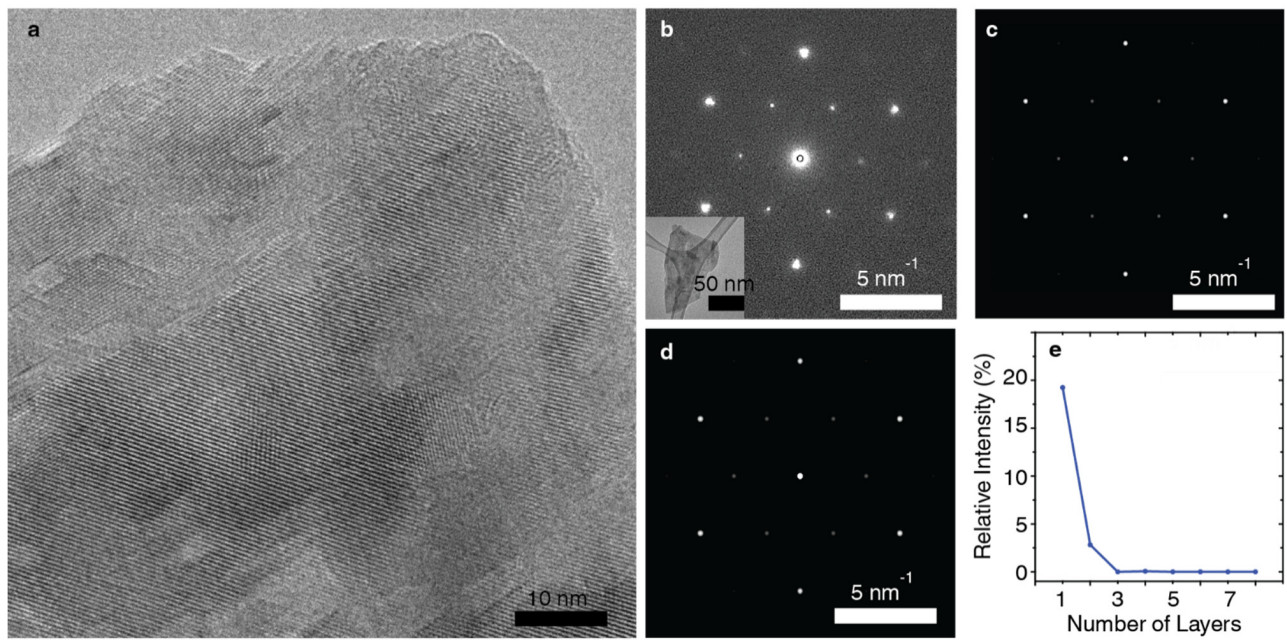

**Figure 2.** Morphology and crystallinity of exfoliated $Ca_2N$. a) High resolution TEM image of 2D $Ca_2N$ showing the sheet-like morphology and crystallinity. b) Diffraction pattern of 2D $Ca_2N$ with the low resolution TEM image of the corresponding 2D flake. Simulated diffraction pattern of c) bilayer $Ca_2N$ and d) $Ca_2N$ with translational disorder looking down the [001] zone axis. e) Comparison of the relative intensity of the spots at 3.21 nm$^{-1}$ to the spots at 5.56 nm$^{-1}$ in the simulated diffraction patterns for different thicknesses of 2D $Ca_2N$.

We analyzed selected area diffraction patterns to understand the crystal structure of the nanomaterials. The diffraction patterns showed a hexagonal crystal structure with a d-spacing of 1.80 ± 0.01 Å, which matches the simulated d-spacing (1.80 Å) for the {1,1,0} family of planes. In addition, we observed a second set of hexagonal diffraction spots with a larger d-spacing (3.12 ± 0.02 Å) not present in the simulated patterns of bulk $Ca_2N$.

To understand the origin of the spots with a larger d-spacing, we used multislice calculations (JEMS)[30] to simulate diffraction patterns for crystals of differing thicknesses and translational disorder. We found that these spots are present in the diffraction pattern of $Ca_2N$ structures that break the unit cell's symmetry in the z-direction. For example, the spots are present in simulated diffraction patterns for monolayer and bilayer structures (Figure 2c, 2e) as well as structures with translational disorder (Figure 2d, S5, see SI for details).

One possible explanation for the presence of translational disorder is turbostratic disorder either present in the 3D parent crystal or introduced in the exfoliation process by grinding the crystal into a powder or by sonicating the powder into 2D flakes. This possibility fits with recent DFT calculations that show that the energy barrier to laterally slide planes of $[Ca_2N]^+$ in a crystal is remarkably low, only ~17 meV.[31] Though we have shown that the layer thickness or translational disorder can explain this data, we note that partial oxidation of the sample surface or a surface reconstruction are also possible explanations. Therefore, we conclude that the crystal structure of our 2D $Ca_2N$ is hexagonal with lattice parameters matching the bulk crystal and that there is aperiodicity in the z-direction.

To understand the composition of our 2D material, we measured the calcium and nitrogen content of samples of bulk (3D) $Ca_2N$ and exfoliated (2D) $Ca_2N$. Details of the assays and an experimental confirmation of the applicability of these assays are given in the supporting information. We measured the stoichiometry of the 3D crystal as $Ca_2N_{1.00±0.01}$ and of the 2D material as $Ca_2N_{0.99±0.01}$ with the error reported as twice the standard deviation. A two-sample t-test with a significance level of 0.05 suggests that the compositions of the 3D and 2D material are not different; however, a one-sample t-test with a significance level of 0.05 suggests that the stoichiometry of the 2D material is nitrogen-deficient or calcium-rich compared to the expected ratio of $Ca_2N$. This result can be explained by a small amount of oxidation on the surface of the 2D material that resulted in the loss of nitrogen as ammonia gas. Therefore, it is likely that the exfoliated material has slightly oxidized during the exfoliation process or subsequent handling.

X-ray photoelectron spectroscopy (XPS) provides insight into the chemical environment of the surface of our exfoliated $Ca_2N$. The calcium 2p core electron spectra (Figure 3a, Table 1) of 2D $Ca_2N$ have a Ca 2p doublet with a spin-orbit splitting of 3.5 eV. The Ca $2p_{3/2}$ peak is centered at 347.5 ± 0.2 eV (Table 1) with a full-width-half-max (FWHM) of 1.9 eV. In addition, the Ca 2p spectra show plasmon loss peaks shifted by 7.9 and 11.4 eV relative to the Ca $2p_{3/2}$ center. These values are in agreement with our 3D $Ca_2N$ (Table 1) and with previous literature.[32] After the measurement, we exposed the 2D $Ca_2N$ samples to ambient conditions, which caused oxidation and a corresponding color change from black to white. The XPS spectra of these deliberately oxidized samples match that of $Ca(OH)_2$ (Figure 3a, S6, Table 1). We conclude that the surface of our as-synthesized 2D $Ca_2N$ has not oxidized to $Ca(OH)_2$. However, because the Ca 2p binding energy is relatively insensitive to the local chemical environment,[33] we do not draw further conclusions about the surface of our 2D $Ca_2N$ from the Ca 2p peak positions.

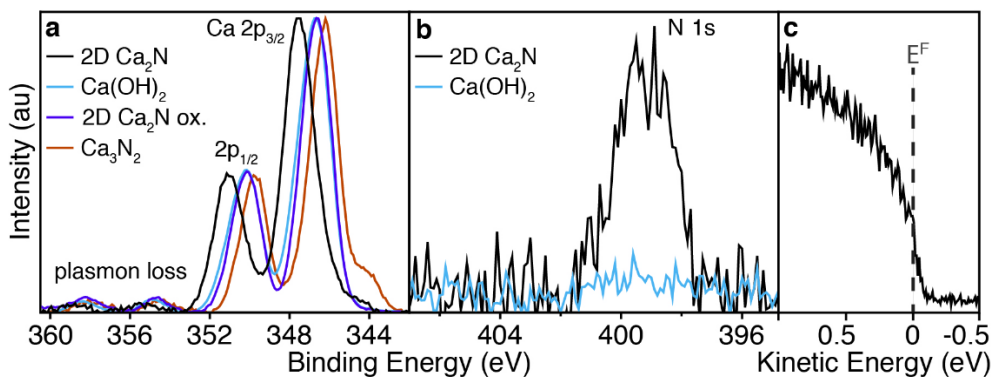

**Figure 3.** Photoemission spectra of calcium nitride species and related oxidized species. a) XPS spectra of core Ca 2p electrons for 2D $Ca_2N$, deliberately oxidized 2D $Ca_2N$, $Ca_3N_2$, and $Ca(OH)_2$. b) XPS spectra of core N 1s electrons for 2D $Ca_2N$ and $Ca(OH)_2$. c) Ultraviolet photoemission spectroscopy Fermi edge of the metallic 2D $Ca_2N$.

**Table 1.** XPS binding energies (eV) of Ca 2p electrons from $Ca_2N$

| Material | $2p_{3/2}$ | $2p_{1/2}$ | Plasm. A | Plasm. B |
|---|---|---|---|---|
| 2D $Ca_2N$ | 347.5 | 351.0 | 355.4 | 358.9 |
| 3D $Ca_2N$ | 347.5 | 351.0 | 355.5 | 359.0 |
| $Ca_3N_2$ | 346.2 | 349.7 | 354.3 | 358.0 |
| $Ca(OH)_2$ | 346.7 | 350.3 | 354.8 | 358.3 |
| 2D $Ca_2N$ ox. | 346.6 | 350.2 | 354.7 | 358.3 |

To further understand the chemical environment of the surface of our 2D $Ca_2N$, we examined the nitrogen 1s core electron binding energy (Figure 3b). We attribute the peak, centered at 399.2 ± 0.1 eV with a FWHM of 2.2 eV, to the N 1s core electrons in 2D $Ca_2N$, in agreement with our 3D $Ca_2N$ (399.3 ± 0.1 eV). We note the N 1s peak is positioned at a different binding energy than previously reported, but we cannot compare absolute positions because a carbon reference was not provided.[32]

The C 1s core electron spectra (Figure S6) show that adventitious carbon species such as adsorbates from 1,3-dioxolane are present on the surface of our 2D $Ca_2N$. While our measurements are representative of the surface of our liquid exfoliated 2D $Ca_2N$, future XPS measurements on clean 2D $Ca_2N$ cleaved in vacuum are needed to provide more insight into the intrinsic bonding and local environment of 2D $Ca_2N$.

Ultraviolet photoelectron spectroscopy (UPS) offers information about the electronic structure of our 2D flakes. The density of states at the Fermi energy, $E_F$, (Figure 3c) demonstrates that the electronic structure of our 2D $Ca_2N$ is metallic (Figure S7). The work function of our material could not be quantified because all samples of bulk and 2D $Ca_2N$ showed evidence of differential charging (Figure S8 and SI for details), in agreement with previous reports.[32]

In order to better understand the electronic structure of our 2D $Ca_2N$ solutions, we measured the optical response of our 2D flakes with UV-visible-near IR (λ = 280-2200 nm) transmission spectroscopy. Dilute solutions of 2D $Ca_2N$ in 1,3-dioxolane were light brown in color and transparent (Figure 4a), with an optical extinction that depended linearly on sample concentration (Figure S9), which allows us to estimate an attenuation coefficient and molar extinction coefficient (Figure 4b). Concentrated solutions (1.05 mg/mL) appear opaque (inset Figure 4b) and transmit less than 1% of light at certain wavelengths.

UV-visible-near IR spectra of 2D $Ca_2N$ suspensions show absorbance peaks at 330 and 480 nm (Figure 4c). To understand the origin of these peaks, we compare our measurements both to our calculated joint density of states (JDOS), for which we used the OptaDOS code,[34-36] and to previously reported[18] experimental data on 3D $Ca_2N$. The JDOS shown in Figure 4d (see SI for details), has local maxima at wavelengths of 360 and 560 nm, in agreement with our experimental results. In addition, we extended a previously reported[18] Drude-Lorentz fit of a reflectivity spectrum for 3D $Ca_2N$ to calculate the attenuation coefficient of the 3D material (Figure 4e, see SI for details). We calculated that the 3D material has local maxima at 360 nm and 520 nm, in agreement with the JDOS and our experimental data of 2D $Ca_2N$. Therefore, we assign the peaks in the UV-vis spectra of 2D $Ca_2N$ to interband transitions.

We can learn more about the nature of the interband transitions by examining the calculated band structure of 2D $Ca_2N$ (Figure 4f). Projections of the electron density integrated over states with energies -1.50 to -1.72 eV (highlighted in blue in Figure 4f) show that the electron density resides in the N $p$ orbitals (Figure 4g), in agreement with previous findings.[19,37] Projections of the electron density integrated over states with energies 0 to +1.00 eV (highlighted in gold in Figure 4f) show that when these states are populated, the electron density occupies the interlayer gap (Figure 4h). Interestingly, direct transitions from the flat band (blue) to the unoccupied conduction band (gold) constitute nearly 100% of the lowest energy transitions, which begin at 900 nm (1.38 eV) to about 700 nm (1.77 eV) (Figure S10). In addition, about 50% of the direct transitions that make up the peak at 560 nm in the JDOS result from transitions from these bands. This suggests that interband absorption events may add electron density to the interstitial electron gas, which could impact the material's work function and electrical conductivity under high intensity illumination.

The near IR data show a response at wavelengths longer than 800 nm (Figure 4b). To qualitatively understand whether the long-wavelength response is dominated by scattering or absorbance, we measured the transmittance with the cuvette inside or outside an integrating sphere (Figure S11). The difference between the spectra, which should be due primarily to light scattering, does not have the same wavelength-dependence as the long-wavelength signal itself. In addition, the attenuation for the two geometries differs by only 20%.

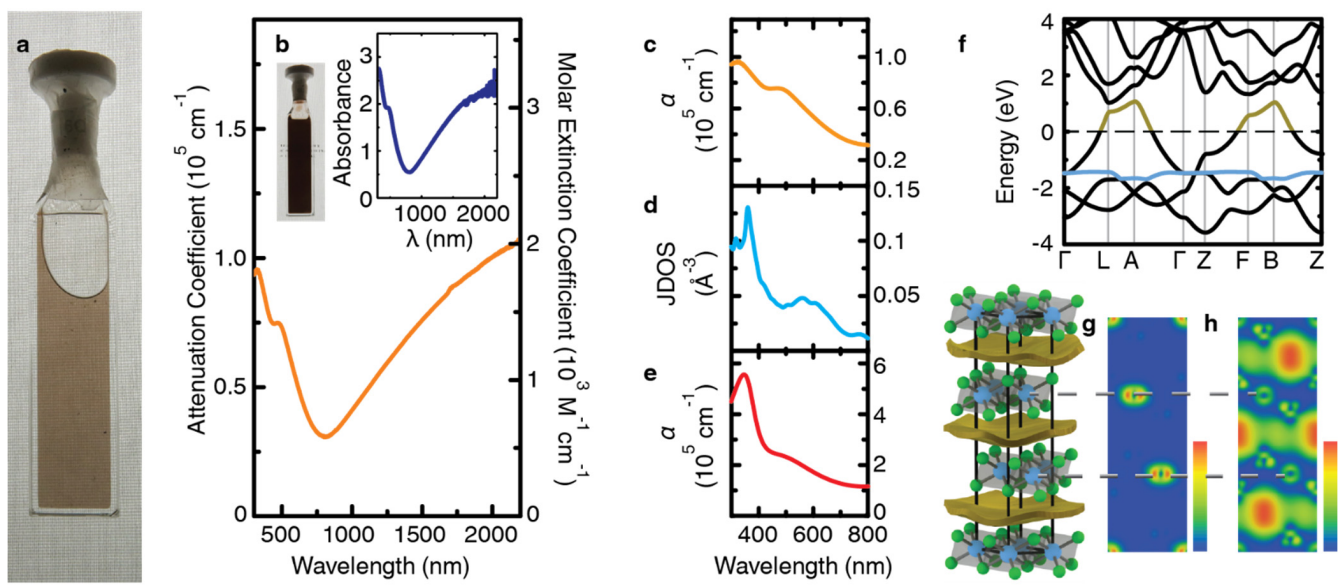

**Figure 4.** Optical properties of 2D $Ca_2N$. a) Photo of a cuvette containing 2D $Ca_2N$ suspended in 1,3-dioxolane in front of the University of North Carolina logo. b) The attenuation coefficient and molar extinction coefficient of 2D $Ca_2N$ vs the wavelength of light. The inset of b) shows a photo of a cuvette containing concentrated 2D $Ca_2N$ and its absorbance spectrum. The short-wavelength response (300-800 nm) comparing c) the attenuation coefficient $\alpha$ of 2D $Ca_2N$, d) the joint density of states calculated by density functional theory for 2D $Ca_2N$, e) the attenuation coefficient $\alpha$ of 3D $Ca_2N$ calculated from literature data.[18] f) The band structure of $Ca_2N$. Projection of the integrated electron density in the unit cell integrated over g) states in the flat band (shown in blue in f) and over states in the band (shown in gold) above the Fermi level. The scale bars for g and h are 0.00 to $1.49 \times 10^{-3}$ e$^-$ Å$^{-3}$ and 0.00 to $0.12 \times 10^{-3}$ e$^-$ Å$^{-3}$, respectively.

This suggests that the near IR response is largely due to the absorbance of light and not scattering. The long-wavelength response is most likely the result of an intraband absorbance of the 2D electron gas, although understanding the nature of the response will require samples that are more uniform in size than are produced in this work (see SI for details).

## CONCLUSIONS

We have demonstrated that layered electrides can be exfoliated into 2D nanosheets despite the electrostatic interactions that hold the layers together. We have shown that the exfoliated 2D flakes are crystalline and metallic, which suggests that the delocalized anionic electrons are preserved in the 2D system.

This work provides the first demonstration of a conceptually new class of high surface-area, electride nanomaterials. These materials combine the high surface area of 2D materials with the exotic properties of anionic electrons. For example, monolayer $Ca_2N$ has a theoretical specific surface area of 1460 m$^2$/g, which far exceeds the highest reported specific surface area (~20 m$^2$/g) of other electrides.[38] At the same time, $Ca_2N$ has a high electrical conductivity ($3.57 \times 10^5$ S/cm),[18] comparable to Al metal,[39] and a low work function (2.4 to 3.5 eV),[17,40] comparable to alkaline earth metals.[41] The material has high transparency for a metal: our experiments show that a 10-nm thick film would transmit 97% of light, while also having a sheet resistance of just 4 $\Omega$/□. The properties of these electride nanomaterials suggest a number of applications, such as reagents or catalysts in chemical synthesis,[5-8] as transparent conductors,[10,11] or as battery electrodes.[12]

## ASSOCIATED CONTENT

**Supporting Information**

3D $Ca_2N$ synthesis and characterization, DFT calculations, Methods of screening solvents for liquid exfoliation of $Ca_2N$, Transmission electron microscopy and simulations, Calcium and nitrogen assays, photoemission, and Optical response of 2D $Ca_2N$.

## AUTHOR INFORMATION

**Corresponding Author**

*sw@unc.edu

**Author Contributions**

‡These authors contributed equally.

**Notes**

The authors declare the following competing financial interest(s): We have filed a patent disclosure on the method of synthesis reported herein.

## ACKNOWLEDGMENT

S.C.W. acknowledges support of this research by UNC Chapel Hill startup funds and NSF grants (DMR-1429407 and DMR-1610861). A.H.W. acknowledges support of this work by the NSF Graduate Research Fellowship under Grant No. DGE-1144081. This work was performed in part at the Chapel Hill Analytical and Nanofabrication Laboratory, CHANL, a member of the North Carolina Research Triangle Nanotechnology Network, RTNN, which is supported by the National Science Foundation, Grant ECCS-1542015, as part of the National Nanotechnology Coordinated Infrastructure, NNCI. The measurements on the optical response of 2D $Ca_2N$ were performed using a Cary 5000 double-beam spectrometer in the UNC EFRC Instrumentation Facility established by the UNC EFRC Center for Solar Fuels, an Energy Frontier Research Center funded by the U.S. Department of Ener-gy, Office of Science, Office of Basic Energy Sciences under Award DE-SC0001011. We thank J.F. Cahoon, A.J.M. Miller, W. You, and group members for supplementary access to furnaces, reagents, and other equipment. We thank A.S. Kumbhar for assistance in HR-TEM imaging. We acknowledge A. Alabanza, T. Davis, and O. Reckeweg for fruitful discussions and M.S. Druffel and L.M. Krull for proof-reading.

# Supporting Information

# Experimental Demonstration of an Electride as a 2D Material

*Daniel L. Druffel,[†] Kaci L. Kuntz,[†] Adam H. Woomer,[†] Francis M. Alcorn,[†] Jun Hu,[†] Carrie L. Donley,[§] and Scott C. Warren\*[†§]*

[†] Department of Chemistry and [§] Department of Applied Physical Sciences, University of North Carolina at Chapel Hill, Chapel Hill, North Carolina 27599, United States.

*Email: sw@unc.edu

**Contents**

## 1. 3D Ca$_2$N Synthesis and Characterization

3D Ca$_2$N was synthesized by the reduction of Ca$_3$N$_2$ (Alfa Aesar, 99%) with Ca metal (Alfa Aesar, redistilled granules ~16 mesh, 99.5%) as adapted from previous literature.[1,2] The Ca$_3$N$_2$ was ground into a very fine powder and Ca granules were added. The mixture was ground lightly together in a 1.02:1 Ca:Ca$_3$N$_2$ molar ratio (total mass of a typical batch: 1.2 g) and pressed into a pellet under ~0.56 GPa of pressure using a hydraulic press. The pellet, along with an additional ~0.600 g of Ca metal, was placed into a pocket of Mo foil (Alfa Aesar 99.95%), which was subsequently crimped closed. The Mo pocket was then sealed inside an evacuated (4 x 10$^{-3}$ mbar) quartz ampoule (18 mm ID, ~6-7 cm in length). The ampoule was heated in a Lindberg Blue tube furnace to 1100 K at a ramp rate of 100K/hr. The temperature was held at 1100 K for 2 days and cooled to room temperature over 24 hours. The additional calcium metal that was added inside of the Mo pocket reacted with the quartz ampoule covering the ampoule wall in shiny black/grey material. We found that adding Ca helped prevent the loss of Ca from the pellet. The obtained Ca$_2$N pellet was black. When broken apart, the material was black and shiny with a blue luster. All materials were stored in a glovebox with a nitrogen atmosphere (oxygen < 0.01 ppm) and all synthetic steps were carried out under nitrogen atmosphere.

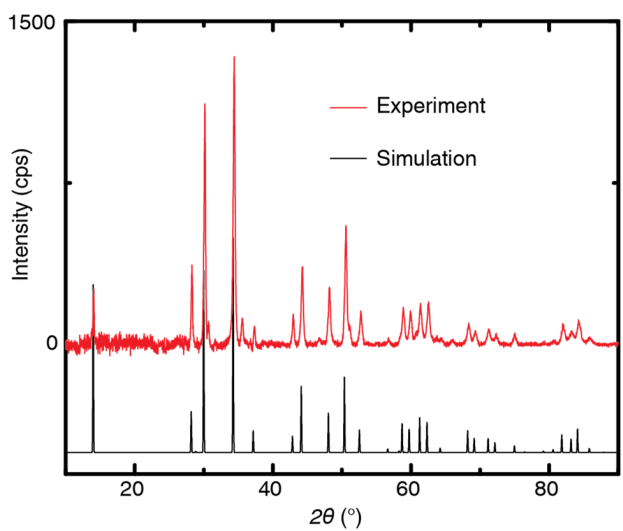

Figure S1. Powder X-ray diffraction pattern of synthesized Ca$_2$N compared to a simulation of the diffraction pattern using reported crystal parameters.[3]

To acquire an X-ray diffraction pattern of Ca$_2$N (Figure S1), the powder was ground very finely, deposited on a roughened glass slide, and covered with 1 mil Kapton tape while in the glovebox to prevent oxidation during the measurement. The measurement was taken using a Rigaku Multiflex X-ray diffractometer with a CuKα X-ray source. We compared our crystal to previous literature[3] and found excellent agreement with the reported hexagonal crystal system, $R\bar{3}m$ space group, and lattice parameters $a$ = 3.624 $c$ = 19.100 Å. The Kapton tape has a broad signal from 10-26°, which reduces the relative intensity of the *003* peak and increases the noise in that region. We note that there are small peaks at 30.6° and 35.7° that we could not assign to a plane in the Ca$_2$N crystal. We attribute those peaks to an unidentified contaminant phase present in many of our samples.

## 2. DFT Calculations of Ca$_2$N and graphite

Density functional theory (DFT) calculations were performed using the CASTEP[4] code with plane-wave basis set[5] approximations. Ultrasoft[6] pseudopotentials were used to describe core electrons, and a 400 eV cut-off energy was used. For this cut-off energy, calculations were convergent with dE/dE$_{cut}$ less than 0.01 meV/atom. A GGA PBE functional[7] was used for the exchange-correlation contribution to total energy and Grimme's DFT-D[8] correction was used to account for long-range dispersion forces. Both graphite and bulk Ca$_2$N were structurally relaxed prior to further calculations. A Monkhorst-Pack[9] grid of 8x8x2 k-points was used for the geometry optimization and interlayer binding energy study. Each crystal structure relaxed to within 2% of experimentally determined parameters.

To determine binding energy for both graphite and Ca$_2$N, we varied the interlayer distance between layers in each crystal structure and then calculated the total energy. An 'infinite' structure was built with an interlayer distance of 12 Å, such that there were no interactions between sheets. We then calculated the interlayer binding energy as either (E-E$_\infty$)/(# of atoms at interface) or (E-E$_\infty$)/(# of interfaces x area).

For electronic structure calculations, we used a denser MK grid, of 64x64x10 and 64x64x64 for the hexagonal and rhombedral orientations of Ca$_2$N. The bandstructure of rhombohedral Ca$_2$N shown in Figure 4e agrees with previously reported calculations[1] and experimentally determined band structures.[10] The integrated electron density was within 0.0001 e$^-$ of the total electron count, and thus we assume all electrons were accounted for and our mesh was sufficient to accurately describe electron density.

The joint density of states (JDOS) was calculated from the CASTEP output with OptaDOS code[11-13] using a Gaussian broadening scheme with a 0.05 eV smearing width. To image the orbital projections of valence and conduction band states, we used the STM profile module of CASTEP for a given negative or positive bias, respectively. Specifically, an applied bias of -1.49 eV was used to image the interlayer electron gas of bulk Ca$_2$N. To obtain the electron density profile, we used Perl scripting to extract and integrate total electron count across the $z$-axis of the hexagonal unit cell.

The band structure of Ca$_2$N changes subtly with thickness (Figure S2a,f) in agreement with other reports[14] and does not change with lateral interlayer translation.[15] Therefore, we expect the band structure of our few-layer flakes to be similar to that of the 3D solid, which was recently shown by angle-resolved photoemission spectroscopy to match the DFT predictions.[10]

Images of the orbital projections for monolayer Ca$_2$N show that both of the bands that cross the Fermi level contribute to the electron density covering the monolayer's surface (Figure S2b,c,d) in agreement with previous findings.[14,16] The electron gas extends as far as 2.0 Å from the calcium atoms of the monolayer. Images of the orbital projections for bilayer Ca$_2$N show that the electron gas sandwiched between layers is at a slightly lower energy than the electron gas at the surface (Figure S2g,j) and that the electron gas at the surface consists largely of filled states in the energy range -0.56 eV to $E_F$ (Figure S2i,j).

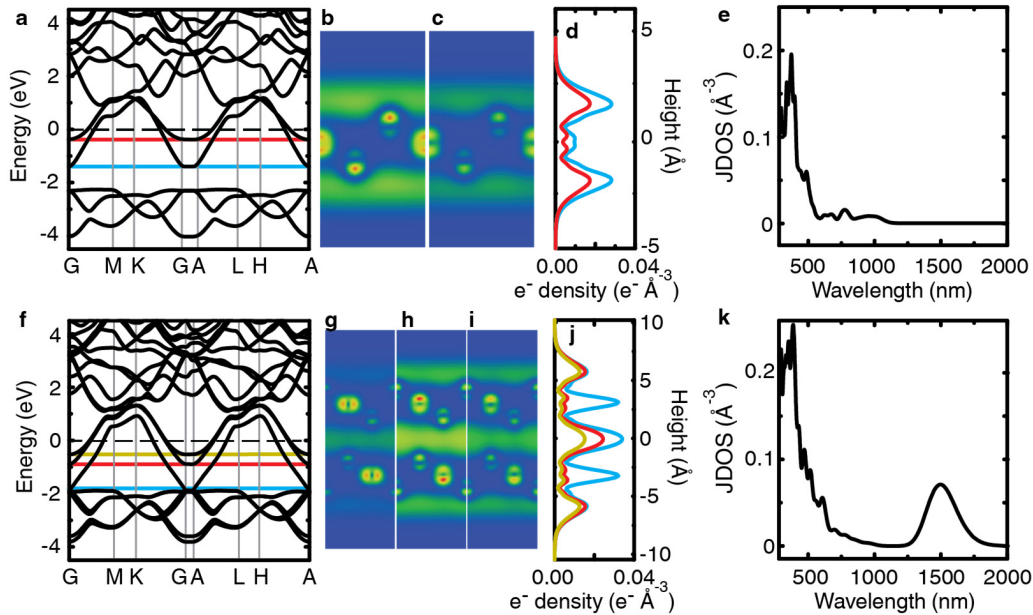

Figure S2. Electronic structure of mono- and bilayer $Ca_2N$. a) Band structure of monolayer $Ca_2N$ calculated from the hexagonal unit cell. Projections of the integrated electron density of monolayer $Ca_2N$ within the unit cell as calculated by density functional theory (DFT). The integration is over states in the energy range b) -1.41 eV (blue line) to $E_F$ and c) -0.39 eV (red line) to $E_F$. The color gradient in b and c ranges from an electron density of 0.00 (blue) to 0.10 x $10^{-3}$ $e^-$ $Å^{-3}$ (red). d) The electron density profile integrated along the z-axis of the unit cell for the states in b and c. e) The joint density of states (JDOS) for monolayer $Ca_2N$. f) Band structure of bilayer $Ca_2N$ calculated from the hexagonal unit cell. Projections of the electron density of bilayer $Ca_2N$ integrated over states in the energy range g) -1.88 eV (blue line) to $E_F$, h) -0.90 eV (red line) to $E_F$, and i) -0.56 eV (yellow line) to $E_F$. The color gradient (blue to red) in g, h, and i varies from 0.00 to 0.41 x $10^{-3}$ $e^-$ $Å^{-3}$, 0.00 to 0.14 x $10^{-3}$ $e^-$ $Å^{-3}$, and 0.00 to 0.14 x $10^{-3}$ $e^-$ $Å^{-3}$, respectively. j) The electron density profile integrated along the z-axis of the unit cell for the states in g, h and i. k) The JDOS for bilayer $Ca_2N$.

The JDOS for the bilayer has a distinct peak at 1500 nm (Figure S2), which arises from direct transitions from the lowest of the three bands crossing the Fermi level to either of the top two bands crossing the Fermi level. These bands are split by ~0.83 eV because of the coupling between the electron layers.[14] The top two bands are primarily composed of states in which the electron density resides on the outer-most electron layer, while the lowest of the three bands is composed of states in which the electron density is sandwiched in between planes of $[Ca_2N]^+$.

### 3. Screening solvents for liquid exfoliation of $Ca_2N$

All solvents were purchased from Sigma Aldrich and dried with 4Å molecular sieves unless otherwise noted. The solvents included: 1,3-dioxolane (anhydrous, 99.5%), dimethyl carbonate (anhydrous, ≥99%), dimethoxy ethane (anhydrous, 99.5%), toluene (Fisher 99.9%), hexane (Fisher 99.9%), benzene (≥ 99.9%), benzyl benzoate (≥ 99.0%), ethyl acetate (Fisher, 99.9%), N-methyl-2-pyrrolidone (anhydrous, 99.5%), 1-vinyl-2-pyrrolidinone (≥ 99%), 1-octyl-2-pyrrolidone (≥ 98%), 1,3-dimethyl-2-imidazolidinone (≥ 99%), N-dodecyl-2-pyrrolidinone (≥ 99%), benzyl ether (≥ 98%), dimethylsulfoxide (≥ 99.9%), chlorobenzene (≥ 99.5%), dichlorobenzene (≥ 99.5%), 1,2,4-trichlorobenzene (≥ 99.5%), cyclohexanone (≥ 99.8%), benzaldehyde (≥ 99%), triethylamine (≥99%), diethyl ether (Fisher, 99.9%), tetrahydrofuran (anhydrous, 99.8%), 1,4-dioxane (anhydrous, 99.8%), dimethylformamide (≥ 99%), dichloromethane (Fisher, 99.9%), acetonitrile (Acros, anhydrous, 99.9%), acetone (Fisher, 99.9%), and *N*-methylformamide (≥ 99%).

Working in a glovebox, we suspended bulk $Ca_2N$ powder (10 mg) in each solvent (20 mL) and sealed the suspensions in polypropylene-capped vials with additional parafilm and Teflon tape wrapped around the lid. We sonicated the sealed suspensions for 100 minutes in a water sonication bath outside of the glovebox with the temperature of the bath below 34 °C.

Table S1. Solvents screened for liquid exfoliation of $Ca_2N$. $Ca_2N$ in compatible solvents (listed in green) remained black after several days and remained suspended. $Ca_2N$ in non-reactive solvents (listed in grey) remained black after several days and precipitated rapidly. $Ca_2N$ in slightly reactive solvents (listed in light yellow) decomposed to a white powder after 5-10 days. $Ca_2N$ in reactive solvents (listed in yellow) decomposed to a white powder after 24-72 hours. $Ca_2N$ in very reactive solvents (listed in red) decomposed to a white powder and produced bubbles within minutes.

| Solvent |
|---|
| 1,3-dioxolane |
| Dimethyl carbonate |
| Dimethoxy ethane |
| Toluene |
| Hexane |
| Benzene |
| Benzyl Benzoate |
| *N*-Methyl-2-pyrrolidone |
| 1-Octyl-2-pyrrolidone |
| *N*-Vinylpyrrolidone |
| 1,3-Dimethyl-2-imidazolidinone |
| *N*-Dodecyl-2-pyrrolidone |
| Ethyl acetate |
| Benzyl ether |
| Dimethyl sulfoxide |
| Chlorobenzene |
| Dichlorobenzene |
| 1,2,4-trichlorobenzene |
| Cyclohexanone |
| Benzaldehyde |
| Triethylamine |
| Diethyl ether |
| Tetrahydrofuran |
| 1,4-dioxane |
| Dimethylformamide |
| Dichloromethane |
| Acetonitrile |
| Chloroform |
| Acetone |
| *N*-Methylformamide |

After identifying 1,3-dioxolane as the most promising solvent, we repeated the screening with 1,3-dioxolane, tetrahydrofuran (THF), and 1,4-dioxane, which have similar structures. $Ca_2N$ (10 mg) decomposed to a white powder after 24-72 hours of exposure to THF or 1,4-dioxane. To ensure that the solvents had very minimal water content and to remove stabilizers, we dried each solvent over sodium metal and distilled the dry solvent over a Schlenk line with $N_2$ environment. The repeated screening yielded the same results; $Ca_2N$ (10 mg) decomposed to a white powder after 24-72 hours of exposure to THF or 1,4-dioxane. The X-ray diffraction pattern of the white powder matches that of $Ca(OH)_2$ (Figure S3).

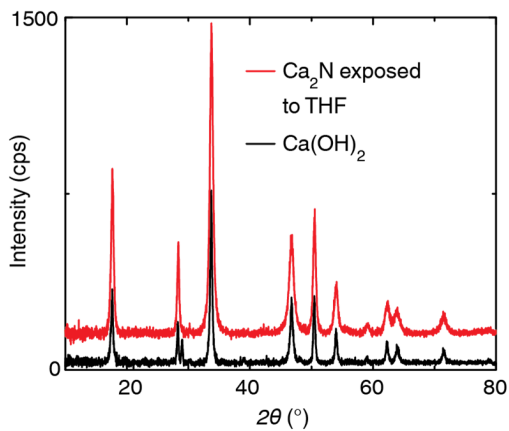

Figure S3. Powder X-ray diffraction pattern of $Ca(OH)_2$ and $Ca_2N$ exposed to THF for 24 hours.

In all subsequent experiments, 1,3-dioxolane was dried over sodium and distilled just before use.

## 4. Transmission Electron Microscopy Experiments and Simulations

In a typical experiment, $Ca_2N$ powder was suspended in 1,3-dioxolane (2.5 mg/mL) as described above and was sonicated for 800 minutes under anhydrous and oxygen-free conditions. The suspension was centrifuged at 300 rpm for 15 min to isolate the exfoliated material. Several microliters of the suspension were drop-cast onto a lacey carbon grid and dried in a vacuum chamber for 30 minutes. The samples were loaded into a high-resolution transmission electron microscope (HR-TEM) using a nitrogen-filled glovebag to minimize exposure to air and water.

To obtain selected area diffraction patterns, the samples were loaded into a low resolution TEM following the same procedure as above. TEM images show that the materials are thin and flat though there is a distribution of flake thicknesses and sizes (Figure S4). Some material is non-sheet-like or appears to be an aggregate of sheets despite our attempt to isolate the thinnest materials.

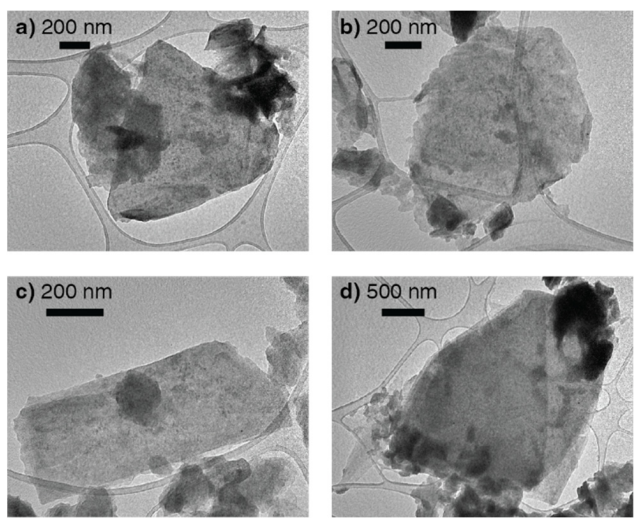

Figure S4. Low-resolution transmission electron microscopy images of flakes of 2D $Ca_2N$.

We simulated the selected area diffraction pattern by multislice calculations (JEMS).[17] To understand the effect of flake thickness on the diffraction pattern, we modeled crystal structures of $Ca_2N$ with thicknesses of 1-8 layers. The monolayer (Figure S5a) shows two sets of hexagonal diffraction spots with d-spacing 1.80 and 3.12 Å, in agreement with our experimental measurements. For flakes thicker than a bilayer, the intensity of the spot corresponding to the larger d-spacing is negligible.

The effect of translational disorder was examined by modeling various stacking configurations of $Ca_2N$. The material is expected to have an ABC stacking sequence in the three layers in the unit cell, but the energetic cost of translating layers is low. We modeled alternate stacking sequences, including ABAC, ABAB, ACAC, ABC-ABC-ABA-ABC-ABC, ACC-ACC-ABA-CBB-BAB. For example, in the sequence ABC-ABC-ABA-ABC-ABC the ninth "C" plane has been replaced with an "A" plane, which is equivalent to translating that "C" plane by 2.09 Å in the direction perpendicular to the X-axis. In the translated structures, like the ABC-ABC-ABA-ABC-ABC stacking sequence (Figure S5c), both sets of diffraction spots are present. The intensity of the spots varies significantly with the stacking sequence. For example, in the highly disordered sequence ACC-ACC-ABA-CBB-BAB (Figure S5d), the diffraction spots of the larger d-spacing (3.12 Å) are more intense than the spots at the smaller d-spacing (1.80 Å).

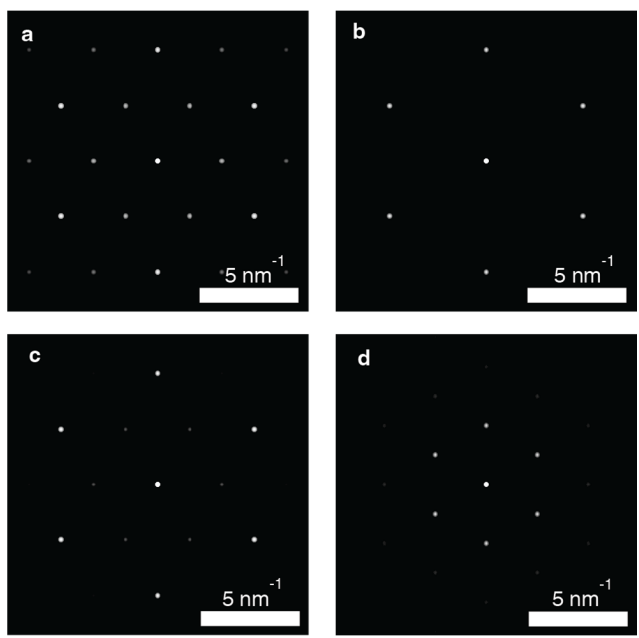

Figure S5. Simulated diffraction pattern of a) monolayer Ca$_2$N, b) trilayer Ca$_2$N, c) Ca$_2$N with stacking sequence ABC-ABC-ABA-ABC-ABC, d) Ca$_2$N with stacking sequence ACC-ACC-ABA-CBB-BAB looking down the [001] zone axis.

## 5. Calcium and Nitrogen Assays

We measured the calcium and nitrogen content of metal nitrides. To determine the concentration of calcium, we titrated Ca$^{2+}$ with ethylenediaminetetraacetic acid (EDTA, Sigma Aldrich, 99%) in the presence of the indicator carconcarboxylic acid (Alfa Aesar, used-as-recieved).[18] First, under nitrogen atmosphere, we digested the samples into (NH$_4$)$_2$SO$_4$ and Ca(OH)$_2$ by injecting 1 M H$_2$SO$_4$ (Fischer, Trace Metal Grade). Then the digested samples were diluted in deionized water to a measurable concentration (about 1.5 mM). The pH of the titrand was kept basic by the addition of NaOH (0.583 M final concentration), which we prepared from NaOH pellets (Fischer, NF/FCC) and deionized water. A stock solution of indicator was prepared fresh for each trial by dissolving carconcarboxylic acid (0.168 mM final concentration) in a 50/50 v/v water-isopropyl alcohol mixture, which was added to the titrand. Sodium potassium tartrate (Sigma Aldrich, ≥99%, 0.0202 M final concentration) was also added to the titrand. The solution was titrated with 0.1202 mM EDTA. Each measurement was repeated ten times.

To measure the amount of nitrogen, we used the Berthelot reaction,[19] a stoichiometric reaction between ammonia and phenol that yields a blue indophenol dye, the concentration of which is quantified by spectroscopy. We digested the samples of metal nitrides into (NH$_4$)$_2$SO$_4$ and Ca(OH)$_2$ by injecting 1 M H$_2$SO$_4$ (Fischer, Trace Metal Grade). Standards (1.860 to 753.2 μM) of (NH$_4$)$_2$SO$_4$ (Alfa Aesar, >99%) with Ca(OH)$_2$ (Fischer, Certified) in a 1:1 ratio were prepared in deionized water and stored in a refrigerator. To the samples and standards, solutions of EDTA (final concentration 0.003479 M), phenol (Alfa Aesar, 99%, unstabilized, 0.06345 M final concentration), sodium nitroprusside (Alfa Aesar, 99%, 0.09375 mM final concentration), $Na_2HPO_4$ (Sigma Aldrich, 99.95%, final concentration 0.9099 mM), NaOCl (Sigma Aldrich, reagent grade, available chlorine 4.00-4.99%, final concentration 0.03885 M), and NaOH (Fischer, NF/FCC, final concentration 0.1610 M) were added. The combined solutions were incubated for 50 minutes to develop color. Then the solution was pipetted into a glass cuvette and measured using a Cary 5000 double-beam spectrometer using 450-800 nm wavelength light. The indophenol dye has a $\lambda_{max}$ = 640 nm. Each measurement was repeated ten times.

We confirm the applicability of this approach to metal nitrides by measuring the stoichiometry of $Ca_3N_2$ (Alfa Aesar, 99%) as $Ca_3N_{2.02\pm0.03}$. In addition, both the calcium and nitrogen assays gave molar concentrations that match the expected amount of sample digested. For example, a sample (19.1 ± 1 mg) of $Ca_3N_2$, which was expected to contain 384 ± 20 µmol of calcium and 256 ± 10 µmol of nitrogen, was measured to contain 384 ± 5 µmol of calcium and 260 ± 4 µmol of nitrogen. Therefore, in addition to an accurate stoichiometry, this method can be used to measure the concentration of unknown masses of metal nitrides, which we used to measure the concentration of our 2D $Ca_2N$ suspended in 1,3-dioxolane.

## 6. Photoemission

X-ray photoemission spectroscopy (XPS) was used to investigate the surface of 2D $Ca_2N$, 3D $Ca_2N$, $Ca_3N_2$, oxidized $Ca_2N$, and $Ca(OH)_2$. Suspensions of 2D $Ca_2N$ in 1,3-dioxolane were drop-cast onto a p-doped silicon wafer with a thick thermal oxide (300nm) under inert atmosphere ($N_2$, <0.01ppm $O_2$, <0.01ppm $H_2O$). The wafer was heated to 75 °C for 15 min and subsequently dried under low vacuum for 15 min to drive off 1,3-dioxolane. $Ca_3N_2$, $Ca(OH)_2$, and oxidized $Ca_2N$ (exposed to ambient conditions for two weeks) were each imbedded into indium foil under inert atmosphere. The samples were then loaded into a Kratos Axis Ultra Delay-Line Detector (DLD) spectrometer under dry $N_2$ conditions and held under a high vacuum (<$10^{-9}$ torr) for analysis. The oxide species were loaded into the XPS instrument separately from air-sensitive samples. The X-ray source was a monochromatic Al Kα source (1487.7 eV). A charge neutralizer (1.8 A filament current, 2.8 V charge balance, 0.8 V filament bias) was used with all oxide species; air-sensitive species were investigated with and without the charge neutralizer. The spectra were corrected to the carbon 1s peak for adventitious carbon (284.6 eV).[20]

By comparing the area of the peaks of the Ca 2p core electrons to N 1s core electrons and accounting for the atomic sensitivity factors, we measured the ratio of Ca:N for 2D $Ca_2N$ (5.3:1), $Ca_3N_2$ (4.6:1), and oxidized $Ca_2N$ (18:1). We found that the ratio is significantly calcium-rich in all measurements and the intensity of the nitrogen peak is low in all samples.

We noticed a trend in the C 1s spectra that the area of the peak at ~289.5 eV increases relative to the area of the peak at 284.6 eV for oxidized 2D $Ca_2N$ relative to unoxidized $Ca_2N$ (Figure S6). This was observed in all samples and may be indicative of $CaCO_3$ or other C=O bond formation.

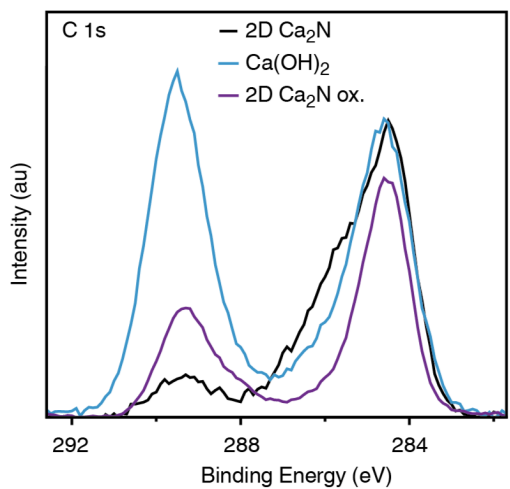

Figure S6. XPS spectra of core C 1s electrons for 2D Ca$_2$N, deliberately oxidized 2D Ca$_2$N, and Ca(OH)$_2$.

Ultraviolet photoemission spectroscopy (UPS) provides information about the valence band of the material and the density of states near the Fermi energy edge. For UPS studies, 2D Ca$_2$N samples were drop-cast onto metal-plated silicon wafers (5 nm adhesive metal, either Cr or Ti, and 50 nm of conductive metal, either Au or Pt). Samples were prepared in the same manner as for XPS analysis described above. A He I source (21.2 eV) was used for UPS measurements.

In order to measure the metallic character of Ca$_2$N and not the substrate, we cast films much thicker than the escape depth of electrons. We characterized the films by XPS and then cleaned the surface of the films by argon ion sputtering. The quality of substrate coverage by Ca$_2$N was assessed by measuring the area of peaks from Pt or Au. We performed sputtering twice for 15 minutes, each time with an accelerating voltage of 1kV and an emission current of 10 mA. The resulting spectra are shown in the Figure S7. After sputtering the surface of 2D Ca$_2$N, we used XPS to determine the relative content of carbon (Fig. S7a), platinum (Fig. S7b), calcium (Fig. S7c), and nitrogen (Fig. S7d). We found that the carbon content decreased as sputtering time increased and that the calcium and nitrogen content increased. This is consistent with the removal of hydrocarbons from the surface of 2D Ca$_2$N. Furthermore, the unchanged position and shape of the calcium and nitrogen peaks after the sputtering indicated that we are not changing or damaging our material during the sputtering process. Additionally, this verifies that the film is generally thicker than the escape depth of the electrons since peaks from the Pt substrate (Fig. S7d) are barely detectable. Furthermore, the density of states up to the Fermi edge (Fig. S7e) becomes more prominent as hydrocarbons are removed from the surface of our thick film. Because XPS probes even deeper into the sample than UPS, we are confident that the density of states at the Fermi edge probed in our UPS measurements are exclusively attributable to the film of 2D Ca$_2$N and not the substrate.

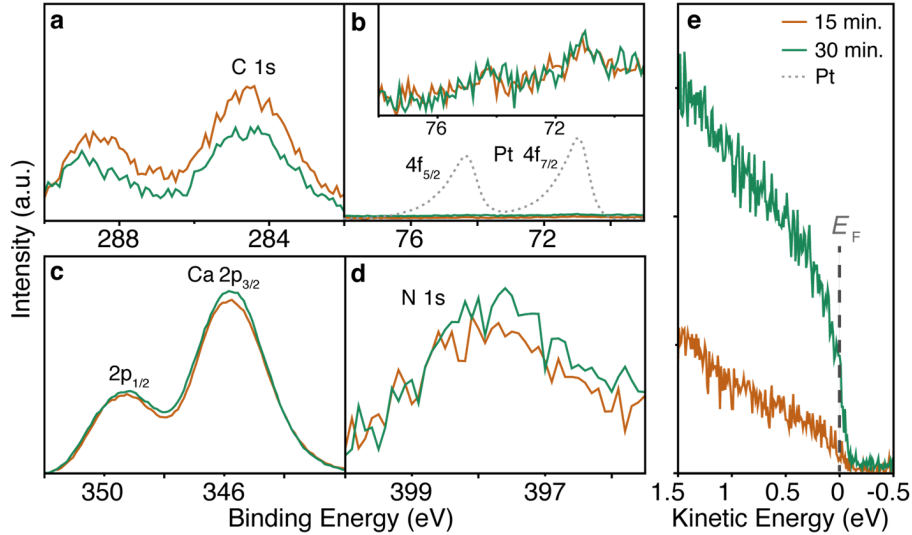

Figure S7. Photoemission spectroscopy of a thick film of 2D Ca$_2$N on a platinum-coated silicon wafer. After sputtering the film with argon ions to remove surface contamination, the film was characterized. In XPS, (a) the carbon content on the surface of 2D Ca$_2$N decreased as sputtering time increased from 15 minutes (orange) to 30 minutes (green). (b) The Pt film is almost entirely covered by Ca$_2$N, as shown by the extremely small peaks at 71.1 and 74.4 eV. The dashed line shows the positions of a bare Pt substrate; the intensities of the dashed line were reduced by a factor of 30. The inset shows a magnified version of the intensities for the Pt core electron binding energies on the film with 2D Ca$_2$N, showing that the Pt signal does not increase with sputtering time. (c) The calcium and (d) nitrogen peaks increase in area as hydrocarbons are removed by sputtering. (e) In UPS measurements, as hydrocarbons are removed, the density of states at the Fermi edge become more prominent and provides evidence for the metallic character of 2D Ca$_2$N.

Generally, UPS allows for calculation of the work function $\phi$ of a material. The work function is related to the Fermi energy of the instrument $E_F$, the energy of the incident photon $h\nu$, and the lowest kinetic energy at which electrons are emitted, $E_{SECO}$:

$$\phi = h\nu - (E_F - E_{SECO})$$

For conductive homogenous samples in electrical contact with the sample holder, the kinetic energy shifts exactly with the biasing potential; however, if different areas of the sample have different electrical conductivity or photoionization cross sections, then the areas of the sample have different steady-state potentials and some features only shift by a fraction of the biasing potential.[21] We find that the surface of Ca$_2$N experiences this effect, which is called differential charging. As shown in Figure S8a, the XPS spectra of Ca 2p core electrons with the charge neutralizer active shows an additional feature that is not present in the spectra when the charge neutralizer is off; this is evidence of differential charging. Similarly in UPS, differential charging affects the secondary electrons emitted and prevents an accurate work function from being measured. The secondary cutoff electron edge $E_{SECO}$ of 2D Ca$_2$N corrected by the applied bias, shown in Figure S8b, have a secondary feature that is induced with increasing applied biases. In Figure S8c, the work functions extracted from Figure S8b are plotted against the applied bias. The slope of the line ($\Delta \phi / \Delta$ bias) varies between samples, but in all cases is much less than 1, which suggests that the sample exhibits differential charging. Unfortunately, this also indicates that we cannot extract a reliable work function from the surface of the material because the energy of the secondary electrons are dependent on the applied bias and the degree of differential charging each sample experiences. We tested other metals (eg. Au, Pt) with known work functions and did not observe differential charging. Therefore, we believe that this response is from the $Ca_2N$ sample and not an artefact of the instrument.

While this response could be caused by poor electrical contact between the sample holder and the drop-cast film, we think that this explanation is unlikely because our observation is consistent over many samples and the same observation was made on 3D $Ca_2N$ in two other studies.[1,22] As suggested by others,[22] the response may be related to the highly anisotropic character of the samples that have areas of markedly different work functions.

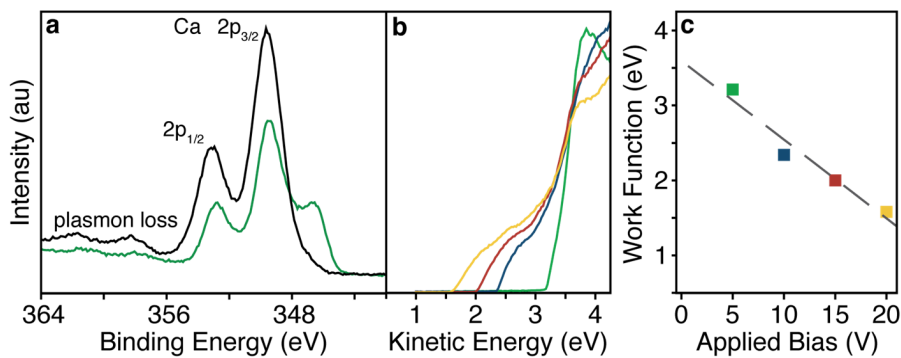

Figure S8. Differential charging of samples of 2D $Ca_2N$. a) XPS spectra of bulk $Ca_2N$ Ca 2p electrons with the charge neutralizer on (green) and with the charge neutralizer off (black). b) UPS spectra of bulk $Ca_2N$ with 5V (green), 10V (blue), 15V (red), and 20V (yellow) applied biases at the secondary cut-off energy, $E_{SECO}$. The spectra are corrected to account for the different applied biases. (c) Work function, $\phi$, plotted against the applied bias for one sample of 2D $Ca_2N$.

## 7. Optical Response of 2D $Ca_2N$

We measured the optical response of 2D $Ca_2N$ with UV-visible-near IR ($\lambda$ = 280-2200 nm) transmission spectroscopy using a Cary 5000 double-beam spectrometer with an external integrating sphere attachment. Quartz cuvettes (Starna 1 mm path length) were filled with suspensions of 2D $Ca_2N$ in 1,3-dioxolane in a glovebox and sealed with parafilm to maintain the $N_2$ atmosphere during the measurement. The attenuation of light through the sample depended linearly on sample concentration across all measured wavelengths (Figure S9).

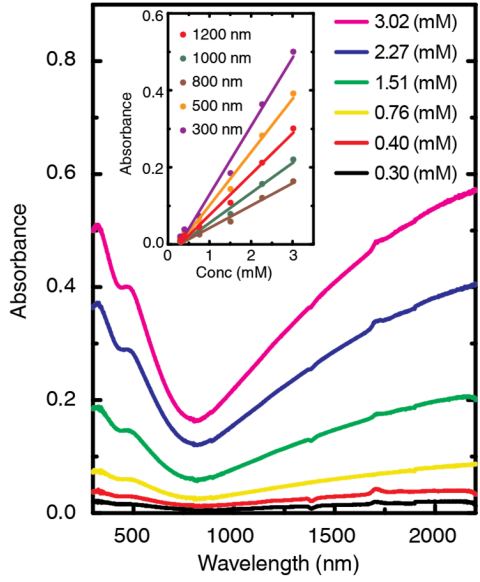

Figure S9. The absorbance of suspensions of 2D $Ca_2N$ vs wavelength. The solvent has been background subtracted. The inset shows that the absorbance depends linearly on the concentration across the UV-Vis-IR range.

Previously reported reflectivity data for 3D $Ca_2N$[1] yielded a fit to the Drude-Lorentz model, which resolved one Drude component, described by a plasma frequency, $\omega_p$, of 2.78 eV and mean scattering time, $\tau$, of 0.64 ps, and two Lorentz components at 2.39 and 3.37 eV.

$$\omega_p^2 = \frac{Ne^2}{m_o \epsilon_o} \tag{1}$$

The Drude susceptibility, $\chi_D$, described by $\omega_p$ and a damping factor $\gamma = \frac{1}{\tau}$ (Equation S2), primarily accounts for intra-band absorbance of free-carriers.

$$\chi_D(\omega) = \frac{-\omega_p^2}{(\omega^2 + i\gamma\omega)} \tag{2}$$

The Lorentz contribution $\chi_{Lj}$, described by an oscillator frequency $\omega_j$ and damping factor $\gamma$ (Equation S3), accounts for inter-band transitions.

$$\chi_{Lj}(\omega) = \frac{\omega_p^2}{(\omega_j^2 - \omega^2 - i\gamma\omega)} \tag{3}$$

The fit can be used to calculate the dielectric function $\epsilon(\omega)$ (Equation S4), which is related to the imaginary part of the refractive index, $k$.

$$\epsilon(\omega) = 1 + \chi_D(\omega) + \sum_{j=1}^{n} \chi_{Lj}(\omega) \tag{4}$$

$$\epsilon(\omega) = \epsilon_1(\omega) + i\epsilon_2(\omega) \quad (5)$$

$$k(\omega) = \frac{1}{\sqrt{2}}\left(-\epsilon_1(\omega) + \sqrt{\epsilon_1(\omega)^2 + \epsilon_2(\omega)^2}\right)^{\frac{1}{2}} \quad (6)$$

$$\alpha = \frac{2k\omega}{c} \quad (7)$$

The attenuation coefficient $\alpha$ (Equation S7), which is reported in Figure 4b, can be calculated by from $k$, the frequency of light $\omega$, and the speed of light $c$.[23]

The fit shows local maxima in attenuation at 360 nm and 520 nm in agreement with the JDOS and our experimental data. The magnitude of the attenuation coefficient predicted by the Drude-Lorentz model only differs from that of our 2D flakes by a factor of three. We note that because the reflectivity data used to make the fit to the Drude-Lorentz model only measured to 350 nm as a high energy bound, the damping term that describes the higher energy Lorentz component is subject to error; as a result, we believe that the attenuation coefficient calculated from this fit is likely overstated at 360 nm.

In addition, we compared our data to the JDOS of $Ca_2N$ calculated from the band structure (Figure S10a). The JDOS accounts for direct transitions from unfilled states to filled states as illustrated by dark blue arrows. Interestingly, transitions from the flat band around -1.7 eV to the unfilled states in the conduction band constitute a large portion of the total JDOS (Figure S10b).

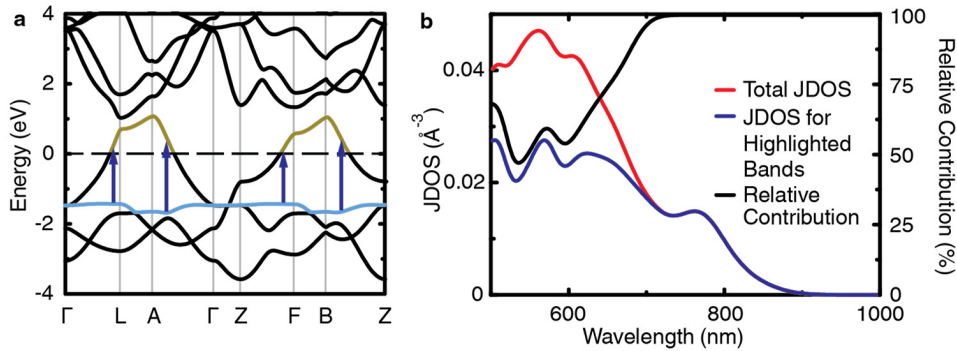

Figure S10. a) The band structure of $Ca_2N$ highlighting the flat band of states with energies of -1.72 eV to -1.5 eV (blue) and the unfilled states in the conduction band with energies 0 eV to 1.00 eV (gold). Dark blue arrows depict direct transitions from the flat band to the conduction band. b) The joint density of states (JDOS) for the entire band structure (red) and for only the transitions from the flat band to the conduction band (dark blue). The relative contribution of those transitions to the total JDOS is given in black.

The experimental near IR data shows attenuation at wavelengths longer than 800 nm (Figure 4b). Because our experiment measures the transmittance of light, the attenuation is a combination of light scattering and absorbance despite our attempts to minimize the scattering component by using an integrating sphere. To try to qualitatively understand whether the long-wavelength response is dominated by scattering or absorbance, we measured the transmittance with 1-cm quartz cuvette inside the integrating sphere and at the front of the integrating sphere (Figure S11). The wavelength was only measured to 1600 nm because the solvent, 1,3-dioxolane, absorbs too strongly in the NIR with a 1-cm cuvette to yield measurable signals.

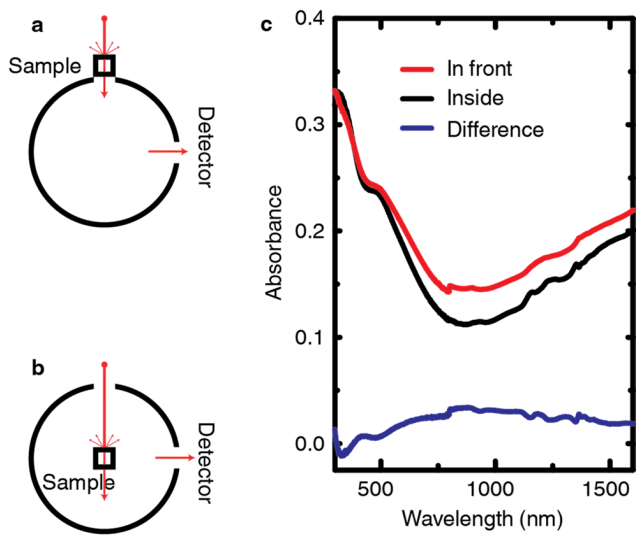

Figure S11. Schematic of the measurement with the sample a) in front of the integrating sphere and b) inside of the integrating sphere. The UV-vis-NIR absorbance spectra of $Ca_2N$ vs wavelength with the sample in different geometries. The solvent has been background subtracted.

The Drude-Lorentz model predicts a minimum in effective attenuation around 800 nm in excellent agreement with our data. In the Drude-Lorentz model, the impinging light is either scattered from the surface or attenuated by electron scattering events inside of the particle. Because our particles have a high surface area to volume ratio, we expect electron surface scattering to decrease the mean scattering time, $\tau$ relative to bulk samples (0.64 ps).[1] Therefore we expect greater attenuation in our 2D $Ca_2N$. Still, a much smaller $\tau$ on the order of 1 fs or smaller would be needed to fit our data to the observed shape and we think that such a small relaxation time is unphysical.

The long-wavelength response could be due to overlapping plasmonic signals from the broad distribution of particle shapes and sizes. However, our measurements seem insensitive to the distribution of particle shapes and sizes. Without controlling for the size and shape of the particle and even with centrifuging at different speeds the long-wavelength response is consistent in all measurements. Therefore, the near IR response could have contributions from Drude absorbance and plasmonic signals. Further studies are needed to understand the long-wavelength response.

## 8. Supporting References